\documentclass{ws-procs975x65}

\newcommand{\be}{\begin{equation}}
\newcommand{\ee}{\end{equation}}
\newcommand{\bel}[1]{\begin{equation}\label{#1}}
\newcommand{\ba}{\begin{eqnarray}}
\newcommand{\ea}{\end{eqnarray}}
\newcommand{\bal}[1]{\begin{eqnarray}\label{#1}}

\begin{document}

\title{UNRAVELING BINARY EVOLUTION FROM GRAVITATIONAL-WAVE SIGNALS AND SOURCE STATISTICS}

\author{I. MANDEL$^*$, V. KALOGERA}
\address{Department of Physics and Astronomy, Northwestern University,\\
Evanston, IL 60208\\
$^*$E-mail: ilyamandel@chgk.info}

\author{R. O'SHAUGHNESSY}
\address{Pennsylvania State University, University Park, PA}

\begin{abstract}

The next generation of ground-based gravitational-wave detectors are likely to observe gravitational waves from the coalescences of compact-objects binaries.    We describe the state of the art for predictions of the rate of compact-binary coalescences and report on initial efforts to develop a framework for converting gravitational-wave observations into improved constraints on astrophysical parameters.

\end{abstract}


\bodymatter

A network of ground-based interferometric gravitational-wave detectors, including LIGO \cite{InitLIGO}, Virgo \cite{Virgo}, and GEO 600 \cite{GEO600}, are currently searching for gravitational waves (GWs) at frequencies between tens and thousands of Hz.  By 2014--5, the Advanced versions of the LIGO and Virgo detectors 
should come online, with sensitivities around ten times greater than for the current network, increasing the volume of the observable GW universe by approximately a factor of a thousand.   Coalescences of compact-object binaries composed of neutron stars (NSs) or stellar-mass black holes (BHs) represent a particularly exciting source for GW astronomy.   Here, we briefly summarize the current astrophysical inputs into GW searches and discuss the potential of GW astronomy to inform our understanding of conventional astrophysics; see Ref.~\refcite{MandelOShaughnessy:2010} for additional details.

Perhaps the most important astrophysical contribution to GW astronomy is the prediction of the rate of detectable GW events, which informs decisions about detector configurations and search techniques.   Compact-object binaries in the field form from isolated primordial main-sequence binaries that evolve through several stages of mass transfer, likely including a common-envelope phase, while the binary components age and eventually undergo supernovae, occasionally leaving behind a tight binary that can merge through gravitational radiation reaction in less than the age of the universe \cite{2007PhR...442...75K}.  In dense stellar environments, such as globular clusters or galactic nuclear clusters, dynamical interactions may contribute a significant dynamical formation rate, particularly for BH-BH binaries (see Ref.~\refcite{MandelOShaughnessy:2010} and references therein).     
Here, we focus on isolated binary evolution and discuss three source populations: NS-NS, NS-BH, and BH-BH binaries.

The merger rate estimates for NS-NS systems can be extrapolated from observations of Galactic binary pulsars via a statistical framework developed in Ref.~\refcite{Kim:2003kkl}.    Five known NS-NS systems will merge in a Hubble time through radiation reaction from GW emission.  In addition to the small number of observations, uncertainties in the extrapolation process primarily come from the need to accurately model selection effects in pulsar searches, which are made difficult by the unknown pulsar luminosity distribution.  The likely extrapolated NS-NS merger rate is $100$ per Myr in the Galaxy, although a rate between $\sim 1$ and $\sim 1000$ per Myr is possible \cite{Kalogera:2004tn}.

There are no observations of compact binaries involving black holes, so the best estimates of the coalescence rate for such systems come from population-synthesis models constrained by electromagnetic observations.   The StarTrack population-synthesis code used in the study in Ref.~\refcite{OShaughnessy:2008} has seven free parameters that can significantly affect the model outcomes: the power-law index in the binary mass ratio; 3 parameters used to describe the supernova kick velocity distribution; the strength of the massive stellar wind; common-envelope efficiency; and the fractional mass loss during non-conservative mass transfer.  Flat priors are assumed on all parameters to sample a wide set of models.  However, several observational constraints, such as the extrapolated numbers of merging and wide Galactic NS-NS binaries, are applied to limit the model space.
For NS-BH binaries, the predicted Galactic merger rate ranges from $0.05$ to $100$ per Myr, with the most likely value at $3$ per Myr \cite{OShaughnessy:2008}.  For BH-BH binaries, the rate ranges from $0.01$ to $30$ per Myr, with a most likely value of $0.4$ per Myr \cite{2007PhR...442...75K}.  Elliptical galaxies, even with a low current rate of star formation, can also contribute to the merger rate through delayed mergers, increasing the overall predicted rates, particularly for BH-BH systems \cite{OShaughnessy:2009}.

Predictions for the rate of mergers and the likely characteristics (masses and spins) of the coalescing binaries can contribute to GW searches by aiding in the selection of the optimal configuration for advanced detectors, and by answering such questions as:  Should searches be expanded to the mass range of intermediate-mass black holes (IMBHs)?  Given limited resources, does it make sense to develop a new search for eccentric binaries?  And what waveforms (e.g., spinning or non-spinning) should be used for existing searches?  

\begin{figure}[htb]
\centering
\includegraphics[keepaspectratio=true, width=0.6\textwidth]{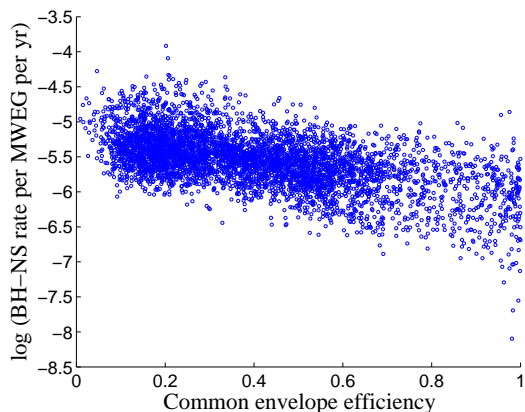}
\vskip-0.1in
\caption{A scatter plot of the predicted BH-NS mergers per Milky Way Equivalent Galaxy vs.~the dimensionless parameter describing the efficiency of the removal of orbital energy by an expelled common envelope (see Ref.~\refcite{OShaughnessy:2005} for definition).}
\label{fig:NSBHvsCEscatter}
\end{figure}

GW astronomy can serve as a new observational window on stellar and binary evolution. Information will come both from individual detections, such as a detection of an IMBH, and from statistics accumulated from multiple observations that can be compared with astrophysical models, like the population-synthesis simulations described above~\cite{OShaughnessy:2008}, to constrain the model parameters.  The rate of detections itself is correlated with some of the model parameters, as indicated in Fig.~\ref{fig:NSBHvsCEscatter} for common-envelope efficiency.  Although this figure suggests that it may be difficult to determine individual model parameters in a highly degenerate parameter space, the detected merger rate can be incorporated as an additional constraint on the allowed model space via a Bayesian framework \cite{MandelOShaughnessy:2010}.  Even in the absence of detections, stringent upper limits can significantly constrain the astrophysical model parameters \cite{MandelOShaughnessy:2010}.

Additional information can come from comparing the distributions of source characteristics, such as masses and spins, with the models.  The tools for accurately estimating these characteristics from the noisy GW signal via Bayesian sampling techniques are largely in hand (see, e.g., Ref.~\refcite{vanderSluys:2008a}).  We recently developed a framework for combining multiple observations into a statement about the distribution of the underlying population \cite{Mandel:2010stat}.  
However, further work is necessary to properly account for selection biases in GW searches, and astrophysical models need to include the full range of theoretical uncertainties in order to allow for meaningful comparisons with observations.  Further developments should enable us to extract the full range of information available in GW data, especially when used in conjunction with possible electromagnetic counterparts, to explore astrophysics and to probe strong-gravity regimes near compact objects.

\section*{Acknowledgments}

The authors acknowledge the support of the National Science Foundation through the Astronomy and Astrophysics Postdoctoral Fellowship under award AST-0901985 (IM), grant PHY-0653321 (VK, IM), grant PHY 06-53462 (RO) and a travel grant that allowed IM to attend MG12.

\bibliographystyle{ws-procs975x65}
\bibliography{Mandel}

\end{document}